\documentclass[manuscript]{aastex}
\usepackage[english]{babel}

\begin{document}

\title{Scaler mode of the Auger Observatory and Sunspots}
\author{Carlos A. Garc\'ia Canal}
\affil{Instituto de F\'{\i}sica La Plata, CCT La Plata, 
CONICET and\\Departamento de F\'{\i}sica, Facultad de Ciencias Exactas,\\
Universidad Nacional de La Plata\\CC 67, 1900 La Plata, Argentina}
\author{Carlos Hojvat}
\affil{Fermilab, P.O. Box 500 -- Batavia IL 60510-0500, USA.}
\author{Tatiana Tarutina}
\affil{Instituto de F\'{\i}sica La Plata, CCT La Plata, 
CONICET and\\Departamento de F\'{\i}sica, Facultad de Ciencias Exactas,\\
Universidad Nacional de La Plata\\CC 67, 1900 La Plata, Argentina}

\begin{abstract} 
Recent data from the Auger Observatory on low energy secondary cosmic ray particles are analyzed to study temporal correlations
together with data on the daily sunspot numbers and neutron monitor data. 
Standard spectral analysis demonstrates that the available data shows
$1/f^{\beta}$ fluctuations with $\beta\approx 1$
in the low frequency range. All data behave like 
Brownian fluctuations in the high frequency range.  
The existence of long-range correlations in the data was confirmed by detrended 
fluctuation analysis. The real data confirmed the correlation between the scaling exponent of the detrended analysis and the 
exponent of the spectral analysis.
\end{abstract}
 

\section{Introduction}

Solar activity gives rise to a modulation
of the flux of cosmic rays observed at Earth. The Pierre Auger 
Observatory \citep{Augergeneral} has made available the scaler singles rates observed
on their surface detectors reflecting the 
counting rates of low energy secondary cosmic ray particles \citep{AugerscalersICRC,Augerscalers,Augerscalers2012}. These data are 
presented after corrections for atmospheric effects,
pressure in particular,
and compared with temporal variations of solar activity as shown by data 
obtained with neutron monitors \citep{neutronmonitors}.
Solar and cosmic rays data can be presented as a temporal series 
containing modulations, correlations and noise fluctuations. The availability of Auger scaler data 
and data from neutron monitors and sunspot numbers \citep{sunspots}
motivated us to study the existence of long-range correlations present in the corresponding time series. 

The Pierre Auger Observatory  is located in the city of Malargüe, 
Mendoza, Argentina. In order to study the highest energy cosmic rays it covers a surface of 3,000 
square kilometers with 1600 surface detectors sensitive to the transversal distributions of the showers 
generated by the primary cosmic rays. In addition, the longitudinal shower development is measured 
by atmospheric fluorescence by 27 optical telescopes. To monitor performance, the surface detectors 
have scalers that record signals received above a given threshold independent of any further shower 
reconstruction involving other neighboring detectors. This is refer to as the ``scaler mode" of the Auger 
surface detector.
The low threshold rates or scaler data have been recorded by the surface detectors of 
Auger Observatory since March 2005.
These data should be sensitive to transient events such as Gamma Ray Bursts and 
solar flares.  The rates at each detector are registered 
every second and the 15 minute average rates are available for public use \citep{Augerscalersonline}.
The temporal variations can be accurately studied, as these rates are very large 
as compared to other data on solar activity.  

In the work of \citet{Augerscalers} the pressure corrected Auger scalers were compared 
to data from the Rome neutron monitor \citep{Rome} and it was concluded that Auger 
scalers could be suitable for the study of solar activity.

A sunspot is a temporary phenomenon in the solar photosphere that appears like a 
dark visible spot compared to the surrounding regions (see, for example, \citet{ssbook}). 
It corresponds to a relatively cool area of the Sun photosphere (1500 K less than the average 
photosphere temperature) as a result of the heat convection process inhibition by intense magnetic fields.

The number of sunspots and their position on 
the Sun face change with time as a consequence of the solar activity cycle. 
The maximum solar activity corresponds to a large number of sunspots and in 
the minimum less sunspots are observed. The spots usually appear in groups. 

Data on sunspots are available for the last four centuries. From 1749 to 
1981 the sunspot data were provided by the Z$\ddot {\rm u}$rich Observatory. 
Nowadays the World Data Center for the Sunspot Index in the Royal Observatory of Belgium  
is responsible for recording sunspot data. All data is available online \citep{sunspots}.

Neutron monitors are ground based detectors that
measure the flux of cosmic rays from the Sun and low-energy  
cosmic rays from elsewhere in the Universe. 
In a typical neutron monitor, low-energy neutrons produced by nuclear 
reactions in lead are slowed down to thermal energies 
by a moderator and detected by proportional counter tubes.
A worldwide network consisting of approximately 50 stations is in operation 
and their data are available on-line \citep{neutronmonitors}.

For this analysis we select data from two neutron monitoring stations. Those  
sites were chosen by the availability of complete data for all the 
days of the period of availability of Auger scaler data. 
Therefore there was no necessity
to perform an interpolation procedure.
These 2 neutron monitor are known as JUNG (\citep{jung}) and APTY 
(\citep{apty}). The 
JUNG detector is located on top of the Sphinx Observatory Jungfraujoch, 
Switzerland and APTY is situated in the town of Apatity, Russia.

The aim of this note is to present a systematic analysis of the 
temporal series from different experimental determinations.
This analysis, based on power spectra behavior \citep{power} and detrended power 
behavior \citep{detrended,power}  
allows us to gain quantitative information about correlations among the 
different phenomena. It is also interesting to study the connection between 
the information provided by the power spectra analysis and the detrended analysis
for the case of real data.

This paper is organized as follows: in section 2 we present the results
of the power spectra analysis of the three data sets and in section 3 we present the 
results of the corresponding detrended analysis. In section 4 final remarks are given.

\section{Power Spectra}

\begin{figure}
\centerline{\includegraphics[scale=0.7]{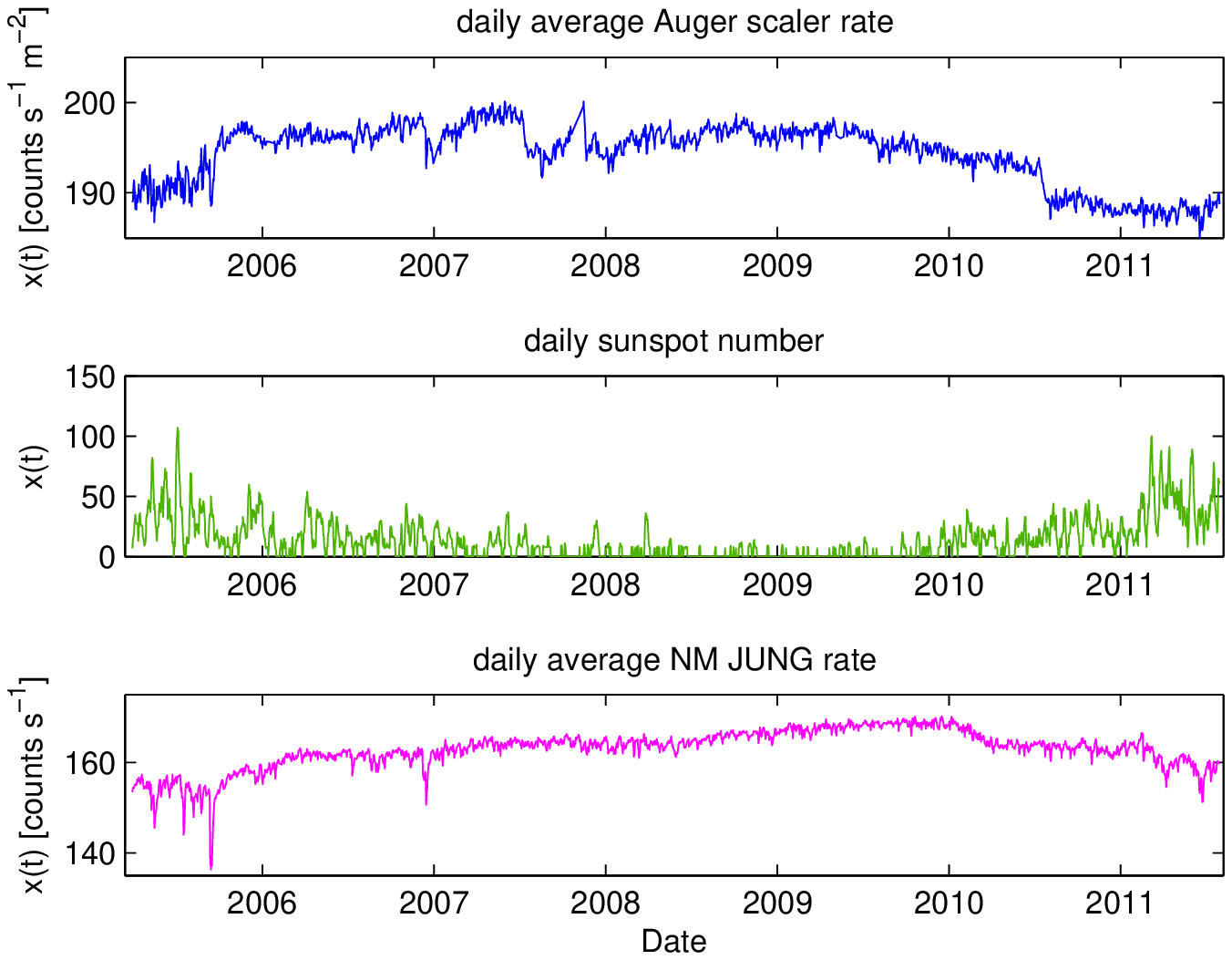}}
\caption{(Color online) Auger scalers, sunspots and neutron monitor JUNG data.}
\label{ts}
\end{figure}

There is a number of accepted techniques \citep{statistics} to characterize 
random processes $x(t)$. Usually one starts with the
correlation function defined as
\begin{equation}
G(\tau)=\langle x(t_0)x(t_0+\tau)\rangle_{t_0}-\langle x(t_0)\rangle_{t_0}^2.
\end{equation}

Another widely used tool is the  frequency spectrum that is defined as the 

squared amplitude of the
Fourier transform of the time signal:
\begin{equation}
S(f)=\lim_{T\rightarrow\infty}\frac{1}{T}\left|\int_{-T}^{T}d\tau x(t) e^{2i\pi 
f\tau}\right|^2.
\end{equation}

For a stationary process, 
the frequency spectrum is connected with the temporal correlation function 
through the Wiener-Khintchine relation \citep{brownian}:

\begin{equation}
S(f)=2\int_0^\infty d\tau G(\tau)\cos(2\pi f\tau).
\label{eq3}
\end{equation}

A true random process, also called white noise has no correlations in 
time, therefore the correlation function for the white noise
is a delta-function and the spectrum function $S(f)\varpropto\rm{const}$. The Brownian motion \citep{brownian}
or random walk corresponds to the spectrum function $S(f)\varpropto 1/f^2$.

Many naturally occurring fluctuations of physical, astronomical, biological, 
economic, traffic and musical quantities exhibit $S(f)\varpropto 1/f$ behavior 
over all measured time scales (see, for example, \citet{Vos75,Pre78, Mat86,Van91,Bai96,Nov97}).
These fluctuations are of interest because they correspond to the
existence of extremely long-range time correlations in the time signals. 
This can be shown (see \citet{Jen98}) if one assumes
 that the spectral function of a time signal $S(f)\varpropto 1/f^{\beta}$ and 
that the temporal correlation function $G(\tau)\varpropto 1/\tau^{\alpha}$. 
It follows from (\ref{eq3}) that 
$1/f^{\beta}\varpropto 1/\tau^{1-\alpha}$. When $\beta\approx 1$ it follows that 
$\alpha\approx 0$ which corresponds to correlation function $G(\tau)\approx 1$.

An analysis of monthly sunspot data was performed in the work of \citet{power}, 
including the study of the power spectra and detrended analysis. It was 
found that the high frequency part of the spectral function of the monthly 
sunspots shows the $1/f^{\beta}$ behavior with $\beta=0.8\pm0.2$. This 
corresponds to the $1/f$ noise.

In Figure \ref{ts} we present the available time series data obtained as indicated in corresponding references: 
(1) Auger scalers \citep{Augerscalersonline}, (2) sunspots \citep{sunspots} and (3) JUNG neutron monitor data \citep{neutronmonitors} for the time period when data on 
the Auger scalers are available. As the Auger scalers 
contained gaps, sometimes of various days, in order to use it in the 
analysis we performed an interpolation of the data justified by the Brownian behavior of 
data for the high frequencies.

\begin{figure}
\centerline{\includegraphics[scale=0.7]{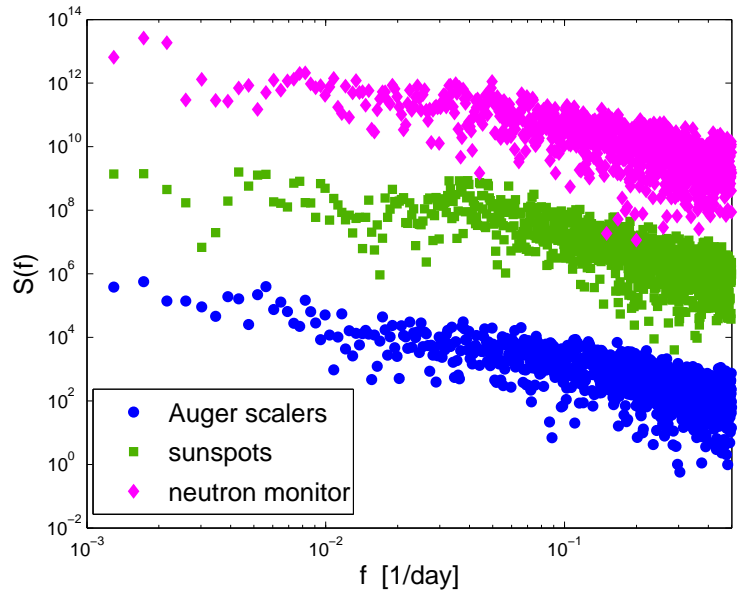}}
\caption{(Color online) Power spectra of daily Auger scalers, sunspots, and  
the neutron monitor JUNG.}
\label{data}
\end{figure}

In Figure \ref{data} we present the power spectra of available data on a logarithmic
 scale vs. frequency. The power spectra for the sunspots
 and the neutron monitor were shifted with respect to each other to avoid 
overlap. 
The spectral function indicates the presence of two different behaviors for low and high frequencies that can be approximated
by linear function with the smaller slope in the range of lower frequencies. The frequency corresponding to the change from one region to another is different for each data set.
We also note that in the sunspot daily spectra there is 
a peak corresponding to the frequency of approximately 1/27=0.04 days$^{-1}$, not clearly seen on Figure \ref{data} because of the low 
statistics. This 27-day peak is due to solar rotation \citep{ssbook}. At first sight the description with the piecewise linear function seems more justified for the spectral function of the sunspots than in the spectral function of the scalers which shows more attenuated behavior for relatively small frequencies. In addition, the slope 
for the low frequency part of the spectrum for the 
Auger scalers and the neutron monitor is smaller compared with that of the sunspots. We make a further analysis of these facts below.

\begin{figure}
\centerline{\includegraphics[scale=0.7]{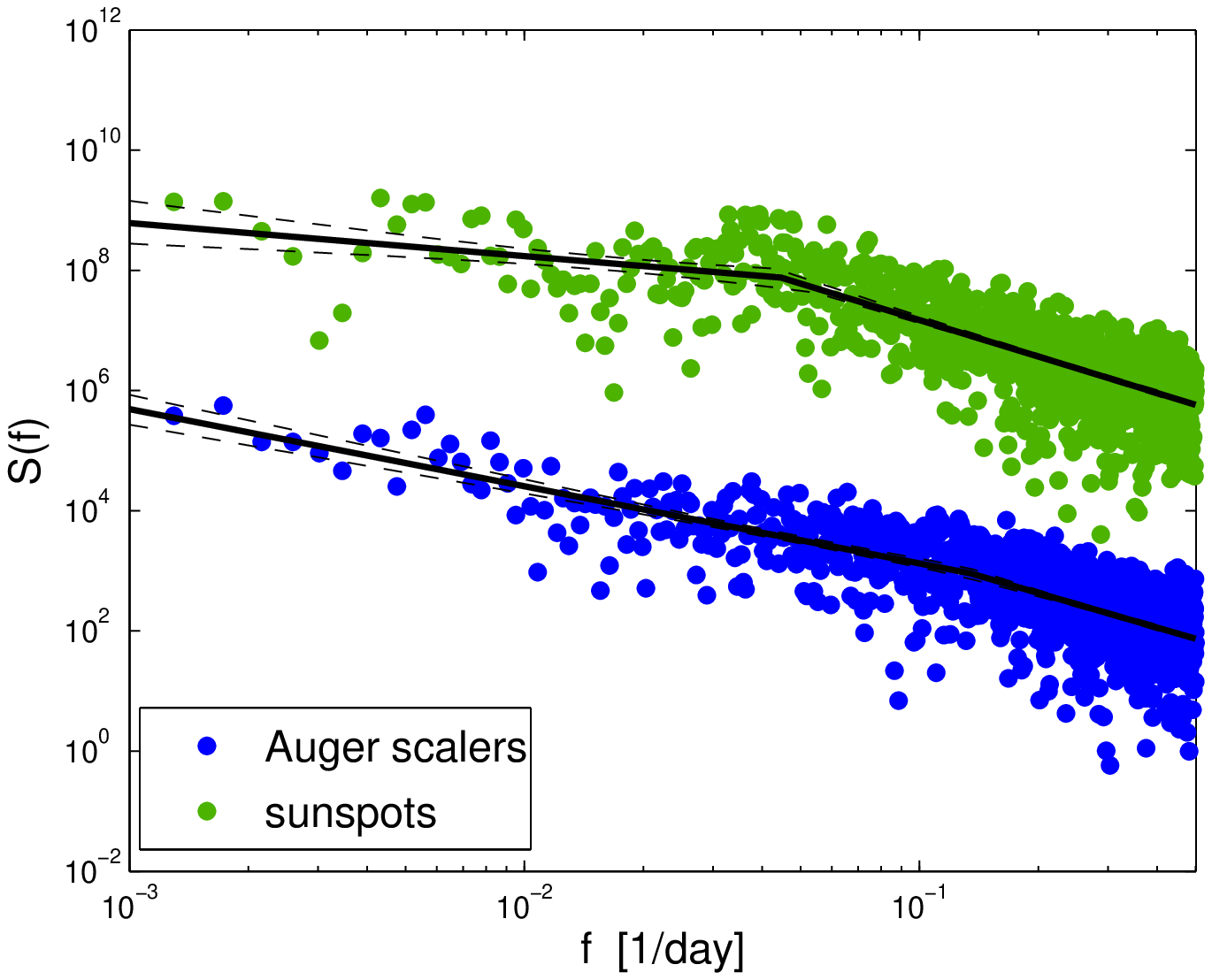}}
\caption{(Color online) Power spectra of daily Auger scalers and sunspots with the corresponding result of the fit with piecewise linear function (solid lines). Dashed lines indicate 95$\%$  prediction bounds.}
\label{fitting1}
\end{figure}

We analyzed the power spectra of Auger scalers, sunspots and two neutron monitors using a linear least squares fit.
First we fitted the data for the total range of frequencies. The resulting $\beta$ is presented in the third line of Table 1.
All data have similar slopes when fitted over the total range of frequencies giving $\beta\approx 1.6$.

Second, we performed a fit of the data to the piecewise linear function
consisting of two parts with different slopes. The frequency corresponding to the point of 
change of the slope $f_c$ was adjusted independently for each data set to obtain the best fit and is presented in the second
line of Table 1. The fitted values of $\beta$ for the range of low frequencies are given in the forth line and for high frequencies -- in the fifth line. In Figure \ref{fitting1} we present the results of the fitting for Auger scalers and sunspots.

It is seen from this analysis that  in the low frequency range all data show the existence 
of long-range correlations with $\beta\approx 1$. For high frequencies, the spectral function shows the Brownian behavior. 
It is also seen that the position of $f_c$ is different for each the data set. 
It should be noted that in the case of sunspots the position of $f_c$ could be
dictated by the existence of 27-day cycle of the Solar activity \citep{ssbook}.

\begin{figure}[ht]
{\centering \begin{tabular}{cc}
\resizebox*{0.5\textwidth}{!}{\includegraphics{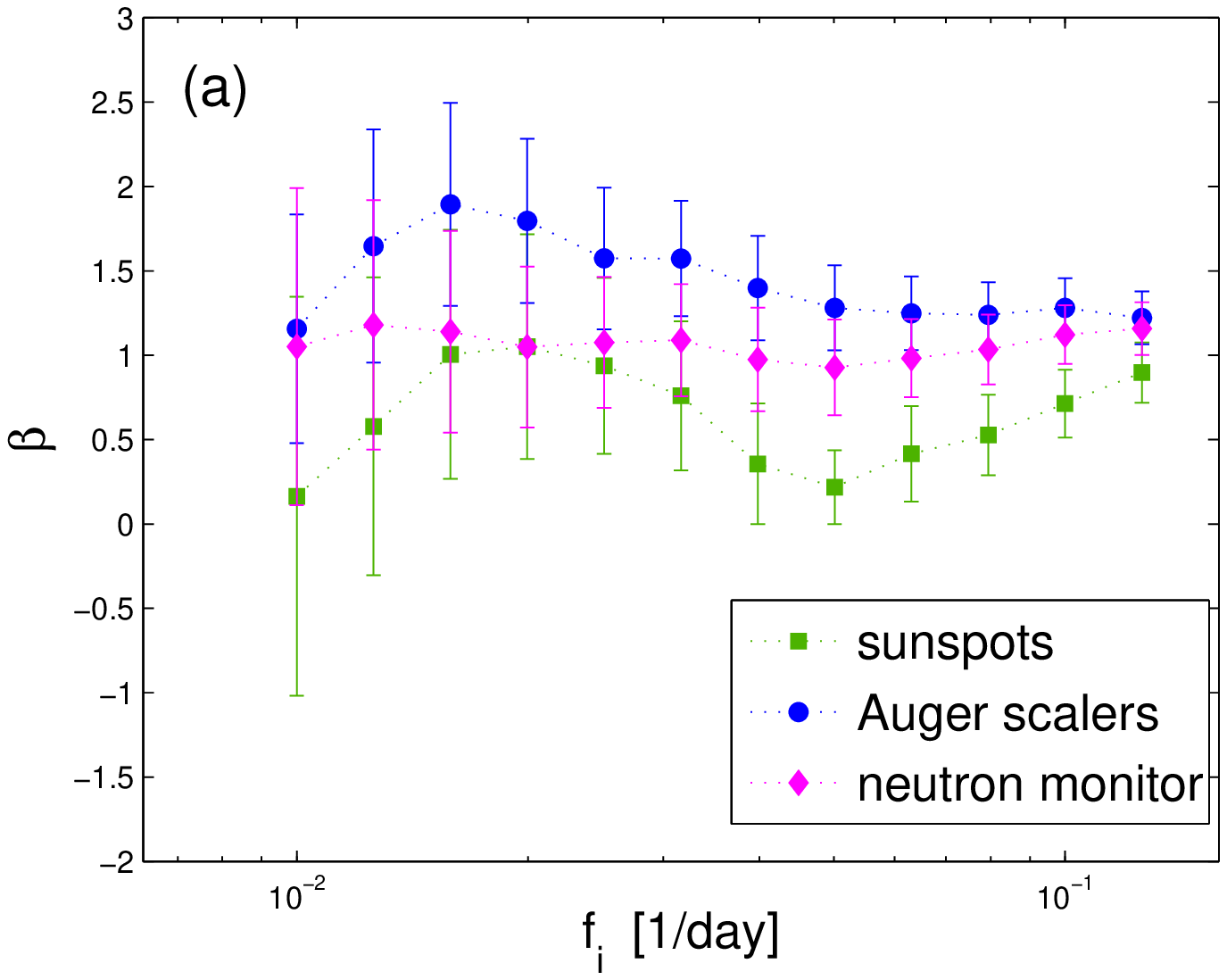}}
&\resizebox*{0.5\textwidth}{!}{\includegraphics{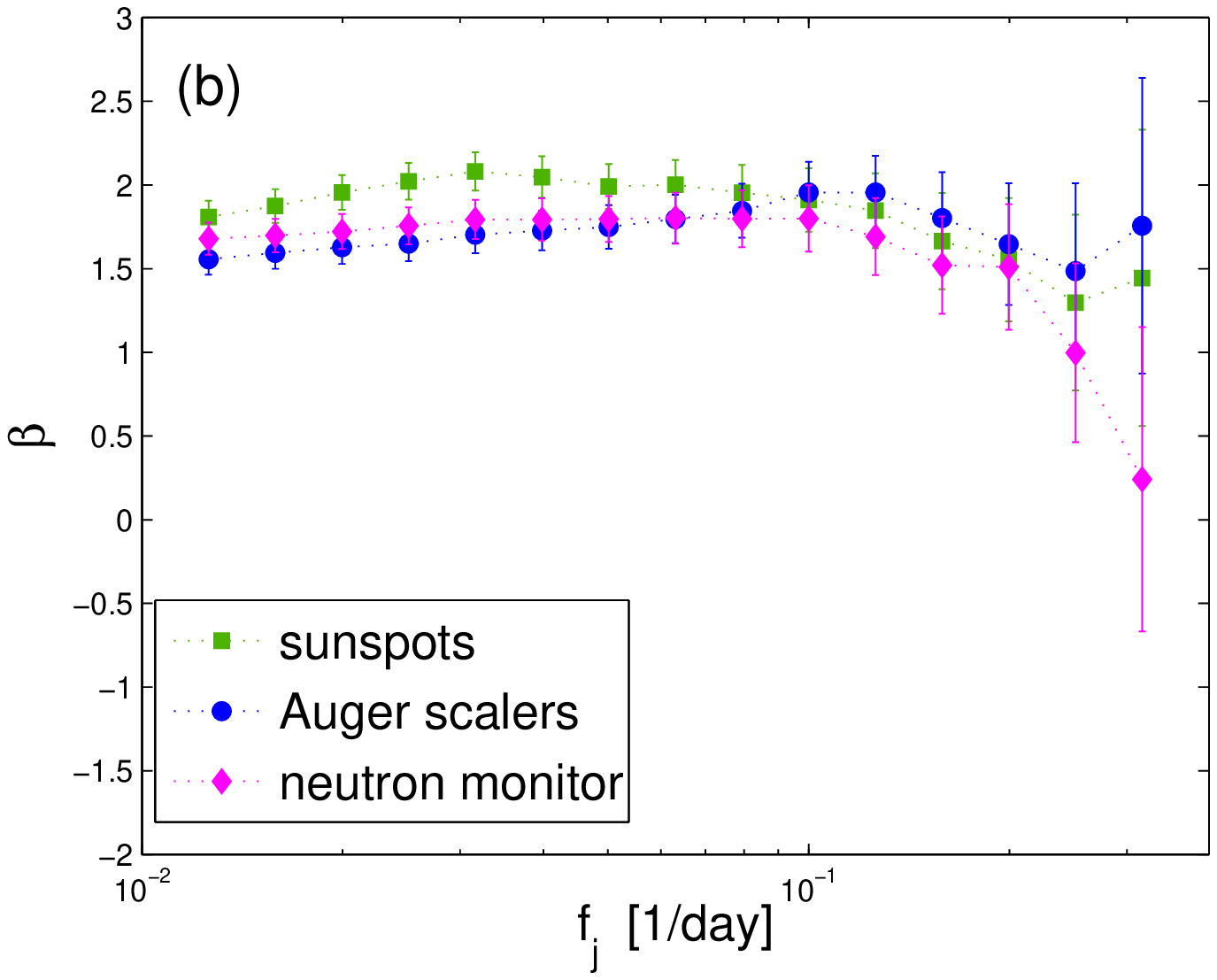}}
\\
\end{tabular}\par}
\caption{The dependence of the power spectra slope $\beta$ on the range of the frequency used for the fit: (a) low frequency range; (b) high frequency range. See text for explanation. The dashed lines are to guide the eye.}
\label{fitting2}
\end{figure}

In order to have more quantitative conclusions about the value of $\beta$  we performed a systematic
fit of all data for different frequency ranges. 
In this analysis we set various sizes of the considered frequency intervals [10$^{-3}$ day$^{-1}$, $f_i$] with $f_i=10^{-3}$ day$^{-1} +\Delta f$ and up to approximately $f_c$ with $\Delta f=10^{0.2}$ day$^{-1}$ and present the fitted slope $\beta$ for sunspots, Auger scalers and neutron monitor JUNG in Figure \ref{fitting2}(a).
For $f_i\approx 10^{-1}$ day$^{-1}$, that is when fitting all sets of data in the range of low frequencies, namely $10^{-3}$ day$^{-1}<f_i<f_c$, $\beta\approx 1$. In the case of sunspots for $f\approx 0.04$ day$^{-1}$ the drop in $\beta$ can be explained by the influence 
of the mentioned 27-day peak which is not seen in the case of Auger scalers and neutron monitors. 
For smaller frequencies the error is large for all the data
because of the low statistics but the results agree within the error bars.

In Figure \ref{fitting2}(b) we show the results of a similar analysis but this time for the  range of high frequencies. Here we obtain the value of
$\beta$ fitting the spectral function over the frequency interval with varying size $[f_j,f_N]$, where $f_j=f_N-\Delta f$ with $\Delta f$ as before, and $f_N$ stands for the Nyquist frequency. It is seen that for $f_j\approx 10^{-1}$ day$^{-1}$ all the data predict a similar value of $\beta$.  As $f_j$ approaches $10^{-2}$ day$^{-1}$ its value drops because of the influence of the part of the low frequency spectra.


 For the high frequency range, the data 
show agreement with the following behavior:
\begin{equation}
S(f)\approx f^{-1.9\pm 0.2} 
\end{equation}
which corresponds to Brownian fluctuations. This is an expected behavior because
 at high frequencies the fluctuations 
become more random or less correlated.

We can conclude that the data show the coexistence of two behaviors in the power spectra:
(1) consistent with a $1/f$ dependence for low frequencies (2) consistent with a $1/f^2$ behavior for high frequencies.
Our analysis confirms the result
of \citet{power} where monthly data on the sunspots was analyzed. The 
sunspots in this frequency region have $1/f$ behavior. The $1/f^2$ behavior
was not detected in \citet{power} because it corresponds to the high frequencies 
which are not present in monthly data.

\begin{table}
\centerline{\begin{tabular}{|c|c|c|c|c|}
\hline
quantity & Auger scalers & Sunspots & NM JUNG & NM APTY\\
\hline
\hline
$\log(f_c/{\rm day}^{-1})$ & -0.87 & -1.35 & -1.25 & -1.15 \\
\hline
$\beta$ total $f$ range & 1.525 $\pm$ 0.073 & 1.591 $\pm$ 0.080 & 1.564 $\pm$ 0.078 & 1.549 $\pm$ 0.079\\
$\beta$ low $f$ & 1.28 $\pm$ 0.14 & 0.55 $\pm$ 0.32 & 1.12 $\pm$ 0.23 & 1.11 $\pm$ 0.22\\
$\beta$ high $f$ & 1.91 $\pm$ 0.23 & 2.02 $\pm$ 0.13 & 1.79 $\pm$ 0.14 & 1.86 $\pm$ 0.16\\
\hline
\end{tabular}}
\caption{Values of $\beta$ for different 
frequency ranges obtained in fitting the power spectra with piecewise linear function with $f_c$ as a free parameter in the fit.}
\label{table2}
\end{table}

\section{Detrended fluctuation analysis}
Another statistical method to reveal the extent of long-range  temporal 
correlations in time series is the Detrended Fluctuation Analysis (DFA).
This method was introduced in \citet{detrended} and applied in many areas of 
research, including physical and biological sciences.  

Consider the time series $x(t)$ consisting of $N$ samples. The procedure is as 
follows:
\begin{itemize}
\item [1.] First generate a new time series $u(t_n)$ by means of: 
\begin{equation}
u(t_n)=\sum_{k=1}^{n}x(t_k),\,\,\, u(0)=0, \,\,\, n=1,\dots, N.
\end{equation}
\item [2.] Divide the time series $u(t)$ 
into non-overlapping  intervals of equal 
length $s$. In each interval the data is fitted using least-squares to the first 
order polynomial (the polynomial of 2nd, 3d order may be used as well, see 
\citet{Kan02}), giving a local trend $y(t)=at+b$.
\item [3.] In each interval calculate the detrended fluctuation function using:
$$
[F(s)]^2=\sum_{t=ks+1}^{(k+1)s}[x(t)-y(t)]^2;\,\,\,\, k=0,1,\cdots,(\frac{N}{s}-1)
$$
\item[4.] Calculate the average of $F(s)$ over $N/s$ intervals.
\end{itemize}

For a fluctuating time series, the expected behavior is as follows:
$$
\langle F(s)\rangle\sim s^{H_{\alpha}}
$$
where $H_{\alpha}$ is a scaling exponent, a generalization of the Hurst exponent \citep{Kan02}. 

Initially  the detrended fluctuation analysis was proposed as an independent 
measure of long-term correlation, complementary to the spectral 
analysis information.
In the work of \citet{Bul95} it was noted that the value of $\beta$ of the spectral 
function and 
$\beta'=2H_{\alpha}-1$ are ``remarkably close to each other''. In this analysis $H_{\alpha}=1/2$ 
corresponds to the case of the white noise, $1/2<H_{\alpha}<1$ reveals the existence of 
positive correlations and the special case when $H_{\alpha}=1$ corresponds to $1/f$ noise. 
Random walk is characterized by $H_{\alpha}=3/2$.
\citet{Hen00} examined the analytical link between DFA and 
spectral analysis and showed that they are related through an integral 
transform. It was concluded that DFA and spectral measures provide equivalent 
characterizations of stochastic signals with long-term correlation. In this work 
we study the connection between the scaling exponent $H_{\alpha}$ and the 
coefficient $\beta$ introduced in the spectral function analysis to 
check this assertion in this particular case with the real data. 

\begin{figure}
\centerline{\includegraphics[scale=0.7]{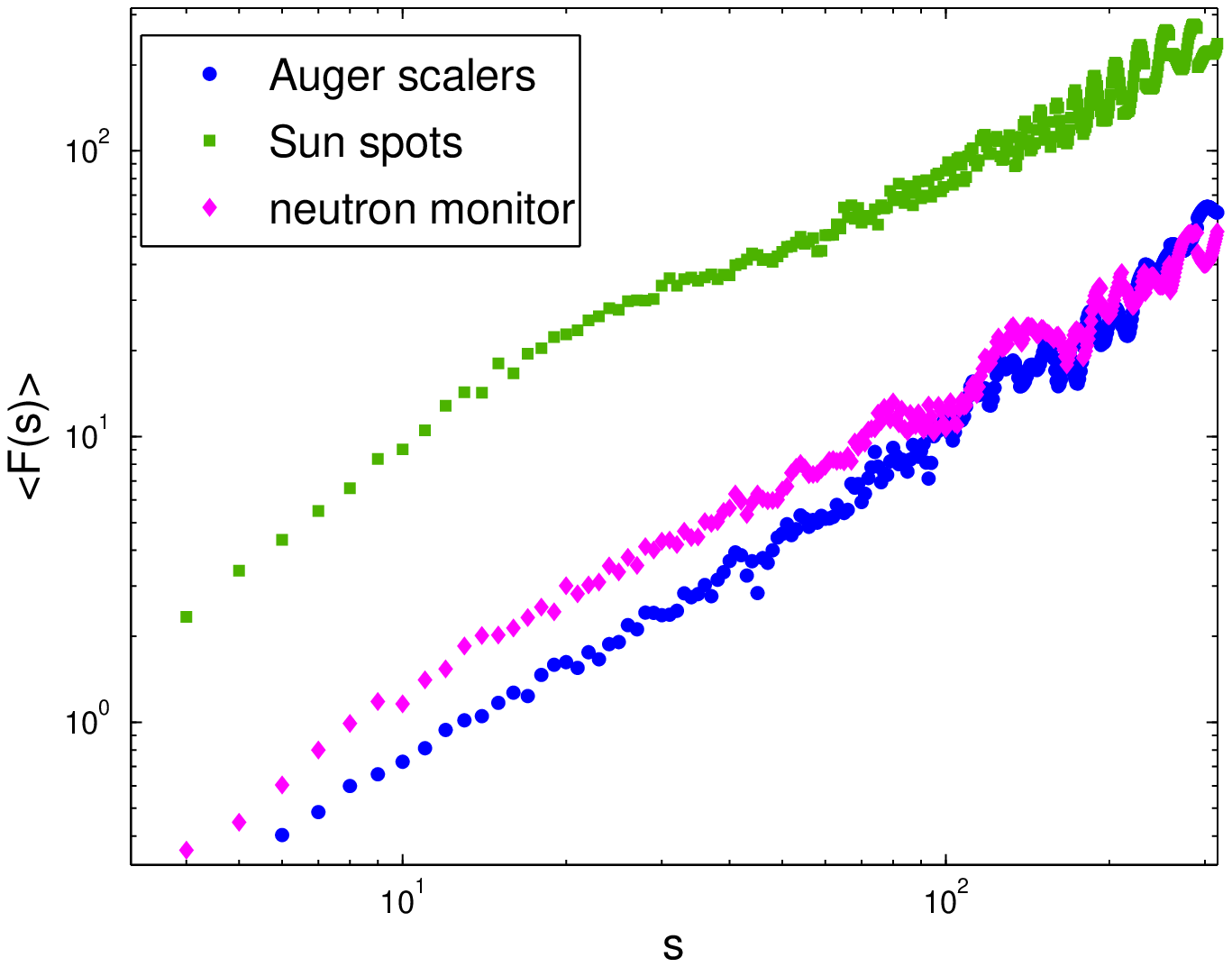}}
\caption{(Color online)Detrended fluctuation analysis curves of Auger scalers, 
sunspots and neutron monitor JUNG.}
\label{fig-det1}
\end{figure}

In Figure \ref{fig-det1} the result of the detrended fluctuation analysis of Auger 
scalers, sunspot numbers and one neutron monitor (JUNG) is shown. In the case of the sunspots, the data can be separated into 3 regions with different slopes: (a) $s\lesssim s_c$; (b) $s_c\lesssim s\lesssim 10^2 $ and 
$s\gtrsim 10^{2}$ (the numbers here are approximate). 
The value of $s_c$ in the case of sunspots approximately 
corresponds to the 27-day peak in the spectral function. For the case of Auger scalers and neutron monitors the existence of three regions in the DFA curve is not as clear as in the case of the sunspots but a close examination of Figure \ref{fig-det1} suggests a change of slope at $s\approx10$ for Auger scalers and neutron monitors. The situation is similar to the power spectra curves where the 
difference between the regions with $1/f$ and $1/f^2$ is more apparent for the case of the sunspots.
There is a correspondence between the frequency in the power spectra of the time series and the value of $s$.
The region of large values of $s$ corresponds to the very small frequencies of the time series and have large fluctuations that can be seen on the Figure \ref{fig-det1}. This region of large values of $s$ is excluded from our analysis.

In Table \ref{table3} we
compare the exponent $\beta$, and $\frac{\beta+1}{2}$, with the 
coefficient $H_{\alpha}$ which characterized the slope obtained in the DFA 
analysis for two ranges of the frequencies: (a) ``high frequency''
$s\lesssim s_c$ and
(b) ``low frequency'' $s_c\lesssim s \lesssim 10^2$, where $\log(s_c)$ is taken equal
to the modulus of the logarithm of the frequency corresponding to the change of the regimen
in the power spectra (and given in Table \ref{table2}).
First of all it is seen that the calculated scaling exponent for low frequencies 
is always $1/2<H_{\alpha}\lesssim 1$ which implies the existence of long-range 
correlations. It should be noted that the analysis of the daily sunspot numbers confirmed the 
result of the analysis of the monthly sunspot number
performed by \citet{power}, where the value obtained for the scaling exponent 
was $H_{\alpha}=0.62\pm 0.4$. It corresponds to the low frequencies 
as the high frequencies are not seen in the monthly data. 

\begin{figure}
\centerline{\includegraphics[scale=0.7]{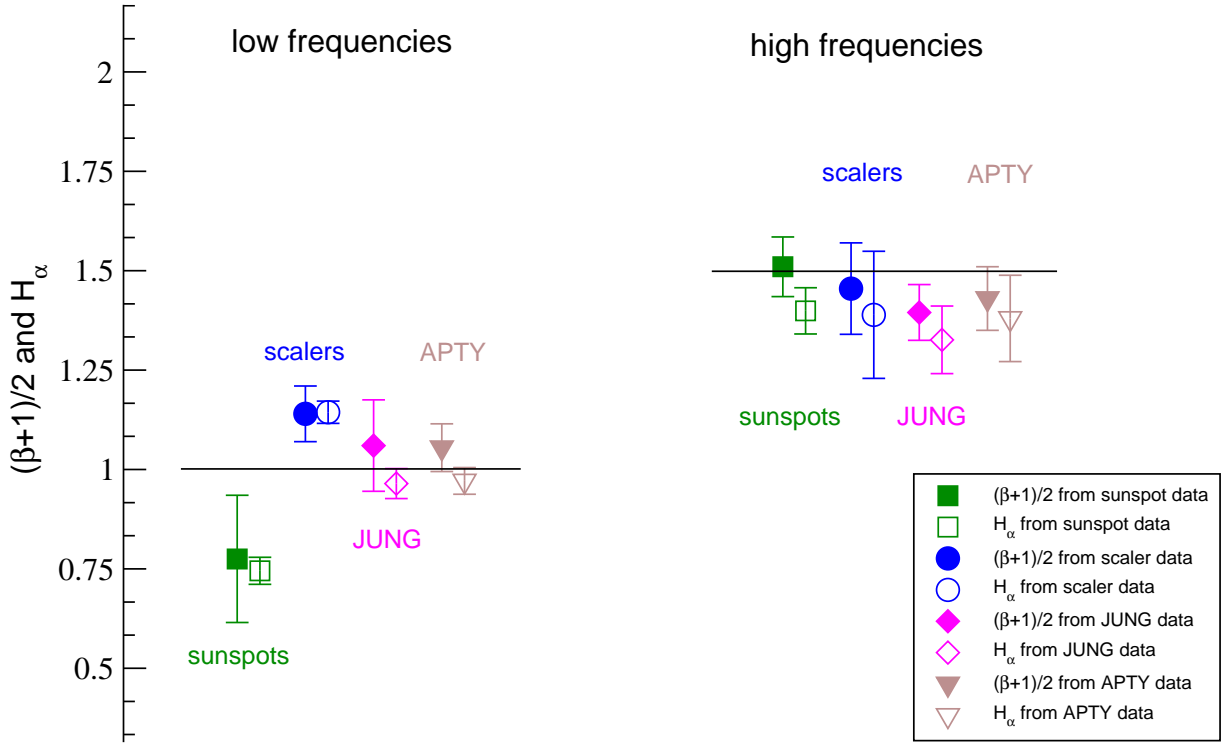}}
\caption{(Color online) Comparison between $\frac{\beta + 1}{2}$ and the scaling exponent for sunspot, Auger scaler and neutron monitor data calculated in the regions of high and low frequencies.}
\label{result1}
\end{figure}

In order to compare the results of power spectra analysis and DFA we present in Figure \ref{result1}
 the slopes calculated using these two methods with the error bars indicated.
The results are grouped by frequency: to the left for low frequencies 
and to the right for high frequencies. In each of these groups we present the resulting 
$\frac{\beta + 1}{2}$ from spectral analysis (filled markers) and the scaling exponent from DFA (open markers).
It is seen that the results of both methods agree within the error bars and are consistent with $H_{\alpha}\approx 1$ for 
low frequencies and $H_{\alpha}\approx 3/2$ for high frequencies. Thus, in this work using real data of three different sources 
we confirm the correspondence between the value of $\beta$ of the spectral 
function and the scaling exponent of detrended fluctuation analysis for both frequency ranges.

\begin{table}
\centerline{\begin{tabular}{|c|c|c|c|}
\hline
frequency & $\beta$ & $\frac{\beta +1}{2}$ & $H_\alpha$\\
\hline
\hline
Sunspots & & & \\
\hline
high & $2.02\pm 0.13$ & $ 1.51\pm 0.08$ & $ 1.40\pm 0.06$\\
low  & $0.55\pm 0.32$ & $0.78\pm 0.16$  & $0.75\pm 0.03$\\
\hline
Auger scalers & & & \\
\hline
high & $1.91\pm 0.23$ & $1.46\pm 0.12$ & $1.39\pm 0.16$\\
low  & $1.28\pm 0.14$ & $1.14\pm 0.07$  & $1.14\pm 0.03$\\
\hline
NM JUNG & & & \\
\hline
high & $1.79\pm 0.14$ & $1.40 \pm 0.07$ & $1.33\pm 0.09$\\
low  & $1.12\pm 0.23$ & $1.06\pm 0.12$  & $0.96\pm 0.04$\\
\hline
NM APTY & & & \\
\hline
high & $1.86\pm 0.16$ & $1.43\pm 0.08$ & $1.38\pm 0.11$\\
low  & $1.11\pm 0.22$ & $1.06\pm 0.06$ & $0.97\pm 0.03$\\
\hline
\hline
\end{tabular}}
\caption{The result of the spectra analysis $\frac{\beta + 1}{2}$ and the scaling exponent $H_{\alpha}$ from DFA for sunspots, Auger scalers and two neutron monitors.}
\label{table3}
\end{table}

\section{Final remarks}
We have presented a statistical study of the available data on 
Auger scalers, sunspots and neutron monitors.
We studied the frequency spectra and performed a detrended analysis.
We found that the spectral function can be separated into two regions: (1) the 
region of the low frequencies 
and (2) the region of high frequencies that can be described by power laws.
It was shown that the low-frequency part of the spectral function of all 
data shows $1/f$ behavior. Because of the low statistics the error is large.
The high frequency part of the spectral function of all data  is shown to 
behave like Brownian fluctuations. This similar behavior of the time series analyzed
can be understood because all three kinds of events are correlated since they have their origin in the solar activity.

The detrended analysis performed confirmed the existence of long-range 
correlations, for low frequencies,  in all available data
(the scaling coefficient $H_{\alpha}\approx 1$). Finally, the correspondence between the scaling 
exponent and the $\beta$ exponent of the spectral analysis
was confirmed with real data.

\section{Acknowledgements}

We would like to warmly thank Professors Huner Fanchiotti and Sergio Sciutto 
for very useful discussions. We acknowledge the Pierre Auger Observatory for 
making the data publicly available and the Pierre Auger Collaboration Publication 
Committee for a critical reading of the text. Fermilab is operated by Fermi 
Research Alliance, LLC under Contract No. De-AC02-07CH11359 with the 
United States Department of Energy. CAGC and TT acknowledge partial support of ANPCyT of Argentina.

\end{document}